\documentstyle[aps,epsf,twocolumn]{revtex}
\newcommand{\be}{\begin{equation}}
\newcommand{\ee}{\end{equation}}

\title{{\bf Measure representation and multifractal analysis of complete genomes}}
\author{Zu-Guo Yu$^{1,2}$\thanks{Corresponding author, e-mail: yuzg@hotmail.com or
 z.yu@qut.edu.au}, Vo Anh$^{1}$ and Ka-Sing Lau$^3$\\
 {\small $^1$Centre in Statistical Science and Industrial Mathematics, Queensland University} \\
 {\small of Technology, GPO Box 2434, Brisbane, Q 4001, Australia.}\\
 {\small $^2$Department of Mathematics, Xiangtan University, Hunan 411105, P. R. China.\thanks{
 Permanent corresponding address of Zu-Guo Yu.}}\\
 {\small $^3$Department of Mathematics, Chinese University of Hong Kong,
 Shatin, Hong Kong}
}

\begin{document}

\maketitle

\begin{abstract}
 {\bf abstract--} This paper introduces the notion of measure representation 
 of DNA sequences. Spectral analysis and multifractal analysis
are then performed on  the measure representations
 of a large number of complete genomes. The main aim of this paper is to discuss
 the multifractal property of the measure representation and the classification
 of bacteria. From the measure
representations and the values of the $D_{q}$ spectra and related $C_{q}$
curves, it is concluded that these complete genomes are not random
sequences. In fact, spectral analyses performed indicate that these
measure representations considered as time series, exhibit strong
long-range correlation. Here the long-range correlation is for the $K$-strings
with the dictionary ordering, and it is different from the base pair correlations
introduced by other people.  For substrings with length $K=8$, the $D_{q}$
spectra of all organisms studied are multifractal-like and sufficiently
smooth for the $C_{q}$ curves to be meaningful. With the decreasing value of 
$K$, the multifractality lessens. The $C_{q}$ curves of all bacteria
resemble a classical phase transition at a critical point. But the
'analogous' phase transitions of chromosomes of non-bacteria organisms  
are different. Apart from
Chromosome 1 of {\it C. elegans}, they exhibit the shape of double-peaked
specific heat function. A classification of genomes of bacteria by assigning 
to each sequence a point in two-dimensional space $(D_{-1},D_1)$ and
in three-dimensional space $(D_{-1},D_1,D_{-2})$ was given. Bacteria that
are close phylogenetically are almost close in the spaces $(D_{-1},D_1)$
and $(D_{-1},D_1,D_{-2})$.\newline
\end{abstract}

\section{Introduction}

DNA sequences are of fundamental importance in understanding living
organisms, since all information of the hereditary and species evolution is
contained in these macromolecules. The DNA sequence is formed by four
different nucleotides, namely adenine ($a$), cytosine ($c$), guanine ($g$)
and thymine ($t$). A large number of these DNA sequences is widely
available in recent times. One of the challenges of DNA sequence analysis is
to determine the patterns in these sequences. It is useful to distinguish
coding from noncoding sequences. Problems related to the classification and
evolution of organisms are also important. A significant contribution in
these studies is to investigate the long-range correlation in DNA sequences$^{
[1-16]}$. Li {\it et al.}$^{\cite{li}}$ found that the spectral density of a
DNA sequence containing mostly introns shows $1/f^{\beta }$ behaviour, which
indicates the presence of long-range correlation when $0<\beta <1$. The
correlation properties of coding and noncoding DNA sequences were first
studied by Peng {\it et al.}$^{\cite{peng}}$ in their fractal landscape or
DNA walk model. The DNA walk$^{\cite{peng}}$ was defined as that the walker steps
``up'' if a pyrimidine ($c$ or $t$) occurs at position $i$ along the DNA
chain, while the walker steps ``down'' if a purine ($a$ or $g$) occurs at
position $i$. Peng {\it et al.}$^{\cite{peng}}$ discovered that there exists
long-range correlation in noncoding DNA sequences while the coding sequences
correspond to a regular random walk. By undertaking a more detailed
analysis, Chatzidimitriou{\it \ et al.}$^{\cite{CDL93}}$ concluded that both
coding and noncoding sequences exhibit long-range correlation. A subsequent
work by Prabhu and Claverie$^{\cite{PC92}}$ also substantially corroborates
these results. If one considers more details by distinguishing $c$ from $t$
in pyrimidine, and $a$ from $g$ in purine (such as two or three-dimensional
DNA walk models$^{\cite{luo}}$ and maps given by Yu and Chen$^{ \cite{YC}}$), then the
presence of base correlation has been found even in coding sequences. On the
other hand, Buldyrev {\it et al.}$^{\cite{Bul95}}$ showed that long-range
correlation appears mainly in noncoding DNA using all the DNA sequences
available. Based on equal-symbol correlation, Voss$^{\cite{voss}}$ showed a
power law behaviour for the sequences studied regardless of the proportion
of intron contents. These studies add to the controversy about the possible
presence of correlation in the entire DNA or only in the noncoding DNA. From
a different angle, fractal analysis is a relative new analytical technique
that has proven useful in revealing complex patterns in natural objects.
Berthelsen {\it et al.}$^{\cite{bgs92}}$ considered the global fractal
dimensions of human DNA sequences treated as pseudorandom walks.

In the above studies, the authors only considered  short or long DNA
segments. Since the first complete genome of the free-living bacterium {\it %
Mycoplasma genitalium} was sequenced in 1995$^{\cite{Fraser}}$, an
ever-growing number of complete genomes has been deposited in public
databases. The availability of complete genomes induces the possibility to
establish some global properties of these sequences. Vieira$^{\cite{Vie99}}$
carried out a low-frequency analysis of the complete DNA of 13 microbial
genomes and showed that their fractal behaviour does not always prevail
through the entire chain and the autocorrelation functions have a rich
variety of behaviours including the presence of anti-persistence. Yu and Wang%
$^{\cite{YW99}}$ proposed a time series model of coding sequences in
complete genomes. For fuller details on the number, size and ordering of
genes along the chromosome, one can refer to Part 5 of Lewin$^{\cite{Lew97}}$.
One may ignore the composition of the four kinds of bases in coding and
noncoding segments and only consider the global structure of the complete
genomes or long DNA sequences. Provata and Almirantis $^{\cite{PY}}$
proposed a fractal Cantor pattern of DNA. They mapped coding segments to
filled regions and noncoding segments to empty regions of a random Cantor
set and then calculated the fractal dimension of this set. They found that
the coding/noncoding partition in DNA sequences of lower organisms is
homogeneous-like, while in the higher eucariotes the partition is fractal.
This result doesn't seem refined enough to distinguish bacteria because the
fractal dimensions of bacteria given by them$^{ \cite{PY}}$ are all the same. The
classification and evolution relationship of bacteria is one of the most
important problems in DNA research. Yu and Anh$^{\cite{YA00}}$ proposed a
time series model based on the global structure of the complete genome and
considered three kinds of length sequences. After calculating the
correlation dimensions and Hurst exponents, it was found that one can get
more information from this model than that of fractal Cantor pattern. Some
results on the classification and evolution relationship of bacteria were
found$^{ \cite{YA00}}$. The correlation property of these length sequences has
been discussed$^{\cite{YAW00}}$.

Although statistical analysis performed directly on DNA sequences has
yielded some success, there has been some indication that this method is not
powerful enough to amplify the difference between a DNA sequence and a
random sequence as well as to distinguish DNA sequences themselves in more
details$^{\cite{hlz98}}$. One needs more powerful global and visual methods.
For this purpose, Hao {\it et al.}$^{\cite{hlz98}}$ proposed a visualisation
method based on counting and coarse-graining the frequency of appearance of
substrings with a given length. They called it the {\it portrait} of an organism.
They found that there exist some fractal patterns in the portraits which are
induced by avoiding and under-represented strings. The fractal dimension of
the limit set of portraits was also discussed$^{ \cite{yhxc99,hxyc99}}$. There are
other graphical methods of sequence patterns, such as chaos game
representation$^{\cite{Jef90,Gold93}}$.

In the portrait representation, Hao {\it et al.}$^{\cite{hlz98}}$ used
squares to represent substrings and discrete colour grades to represent the
frequencies of the substrings in the complete genome. It is difficult to
know the accurate value of the frequencies of the substrings from the
portrait representation. In order to improve it, in this paper we use
subintervals in one-dimensional space to represent substrings and then we
can directly obtain an accurate histogram of the substrings in the complete
genome. We then view the histogram as a measure, which we call
the {\it measure representation }of the complete genome. When the measure
representation is viewed as a time series,  a spectral
analysis can be carried out.

Global calculations neglect the fact that DNA sequences are highly
inhomogeneous. Multifractal analysis is a useful way to characterise the
spatial inhomogeneity of both theoretical and experimental fractal 
patterns$^{\cite{GP83}}$. 
Multifractal analysis was initially proposed
to treat turbulence data. In recent years it has been applied successfully
in many different fields including time series analysis$^{ \cite{pas97,can00}}$
and financial modelling (see Anh {\it et al. }$^{\cite{ATT00}}$). For DNA
sequences, application of the multifractal technique seems rare (we have
found only Berthelsen {\it et al.}$^{\cite{BGR94}})$. In this paper, we pay
more attention to this application. The quantities pertained to spectral and
multifractal analyses of measures are described in
Section 3. Application of the methodology is undertaken in Section 4 on a
number of representative chromosomes. A discussion of the empirical results
and some conclusions are drawn in Section 5, where we also address the use
of the multifractal technology in the classification problem of bacteria.

\section{Measure representation}

\begin{figure}[tbp]
\centerline{\epsfxsize=8cm \epsfbox{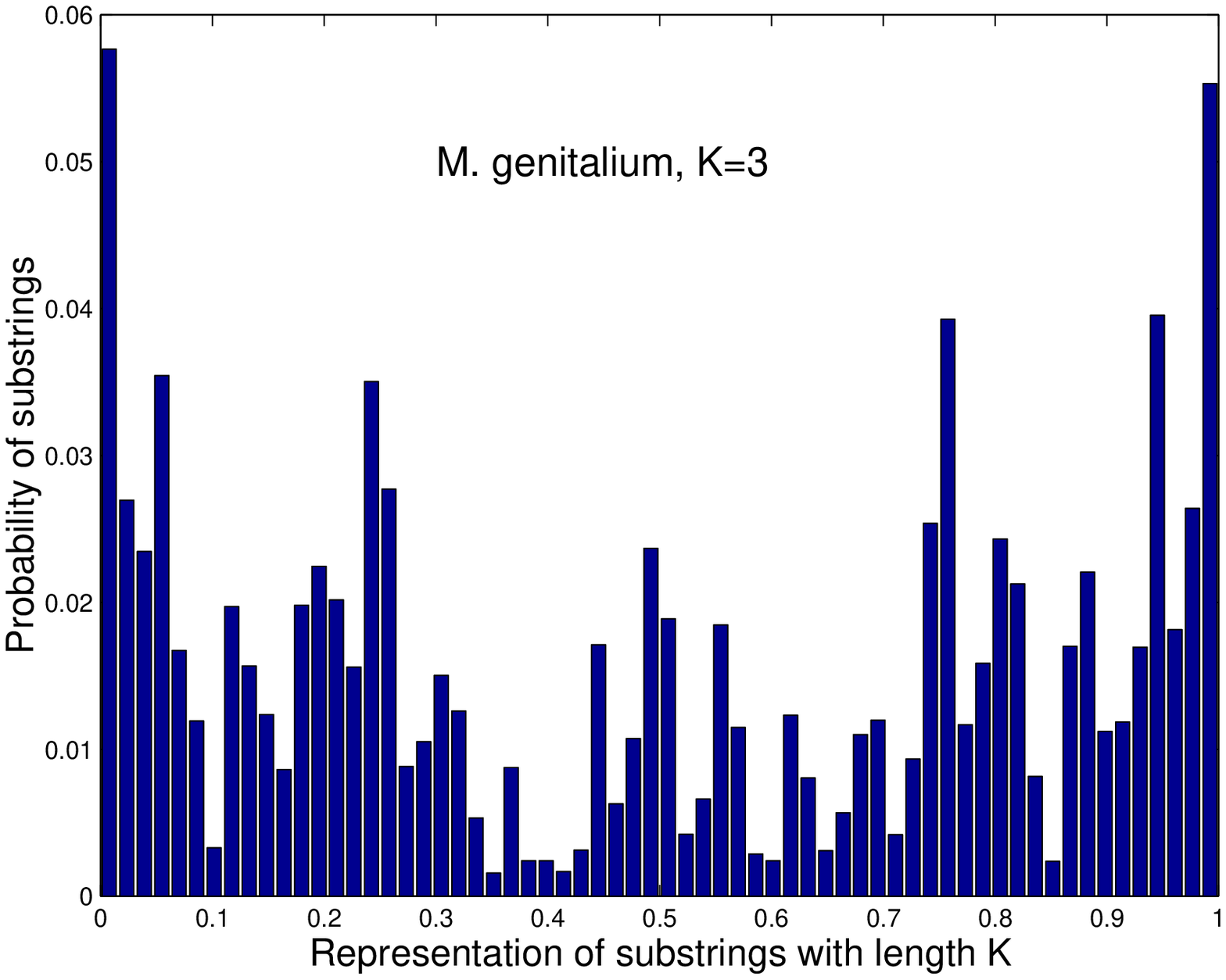}
\epsfxsize=8cm \epsfbox{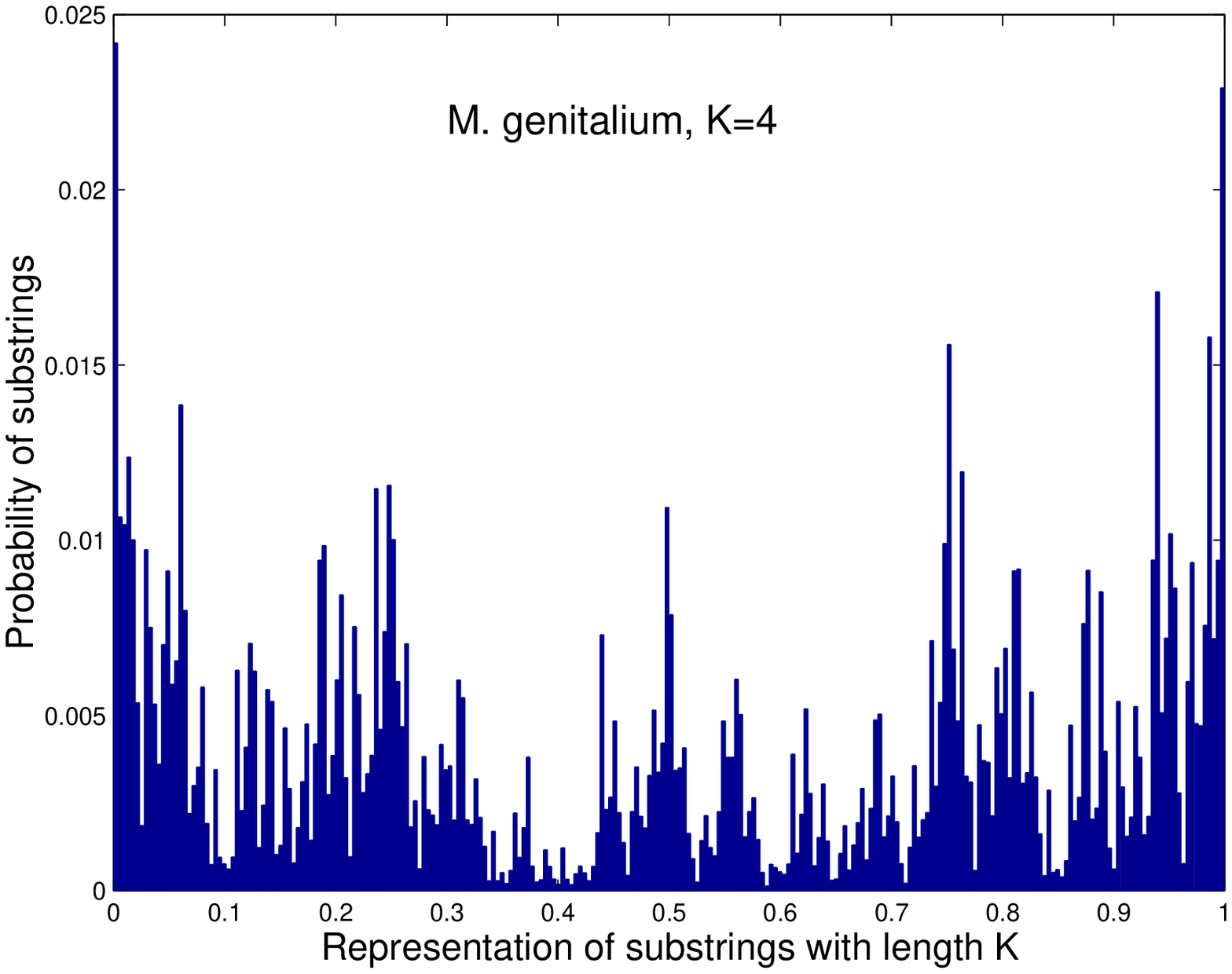}} 
\centerline{\epsfxsize=8cm \epsfbox{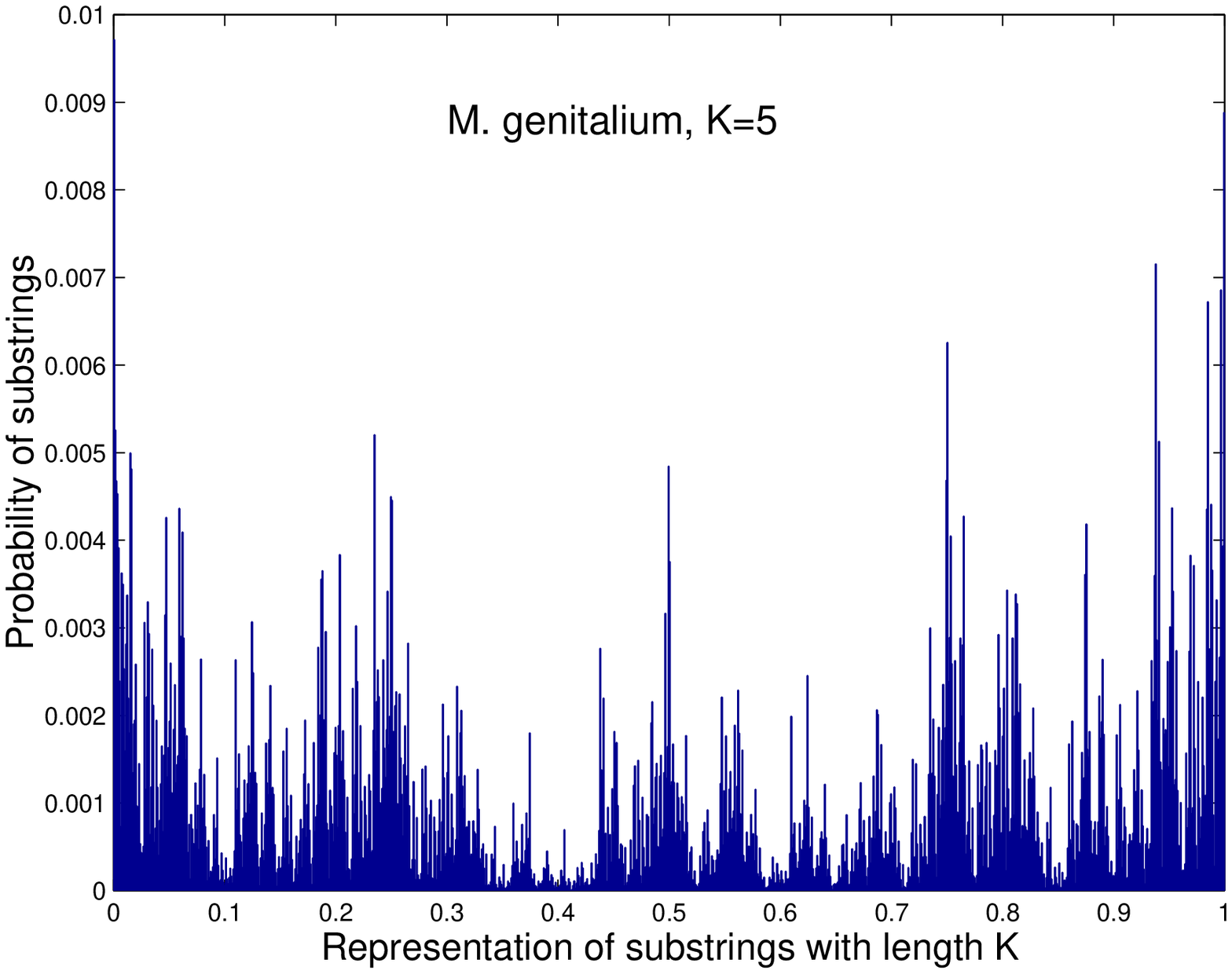}
\epsfxsize=8cm \epsfbox{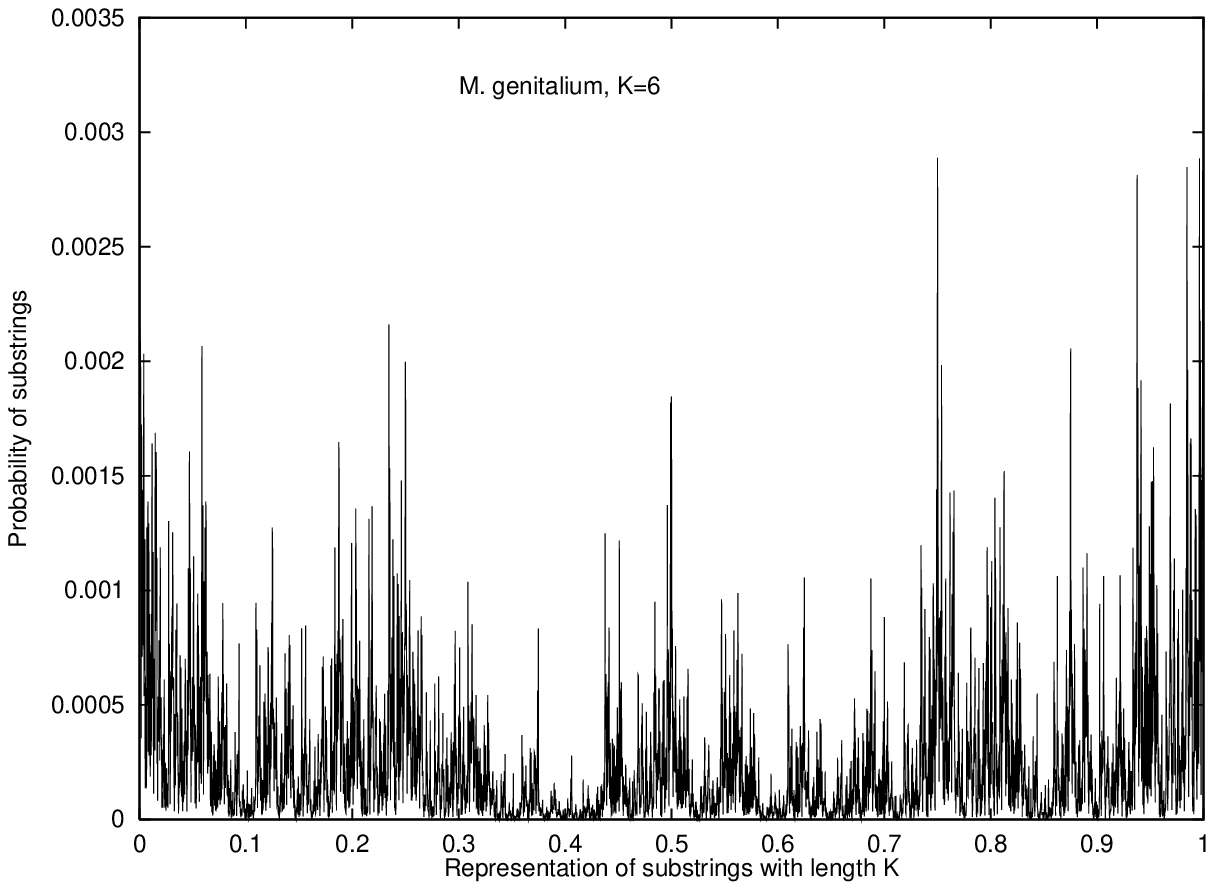}}
\centerline{\epsfxsize=8cm \epsfbox{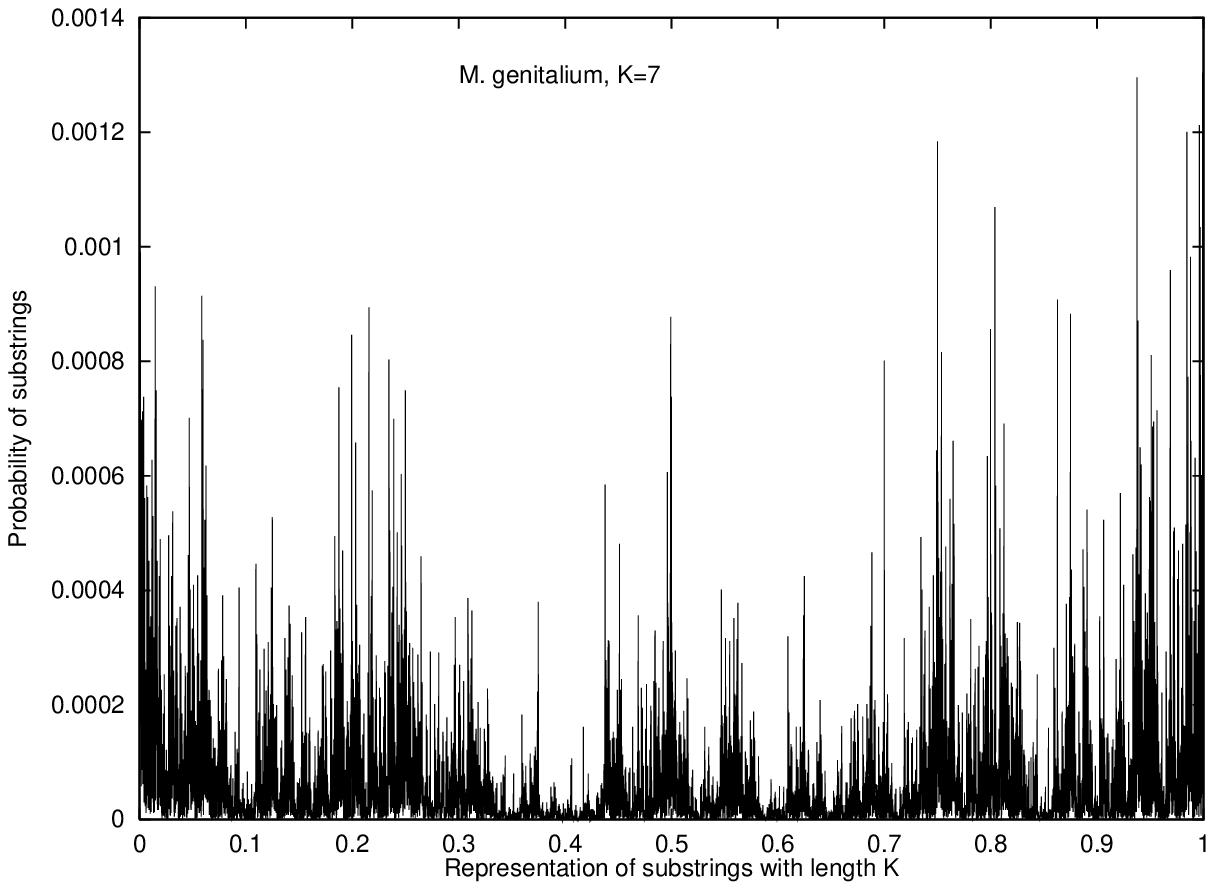}
\epsfxsize=8cm \epsfbox{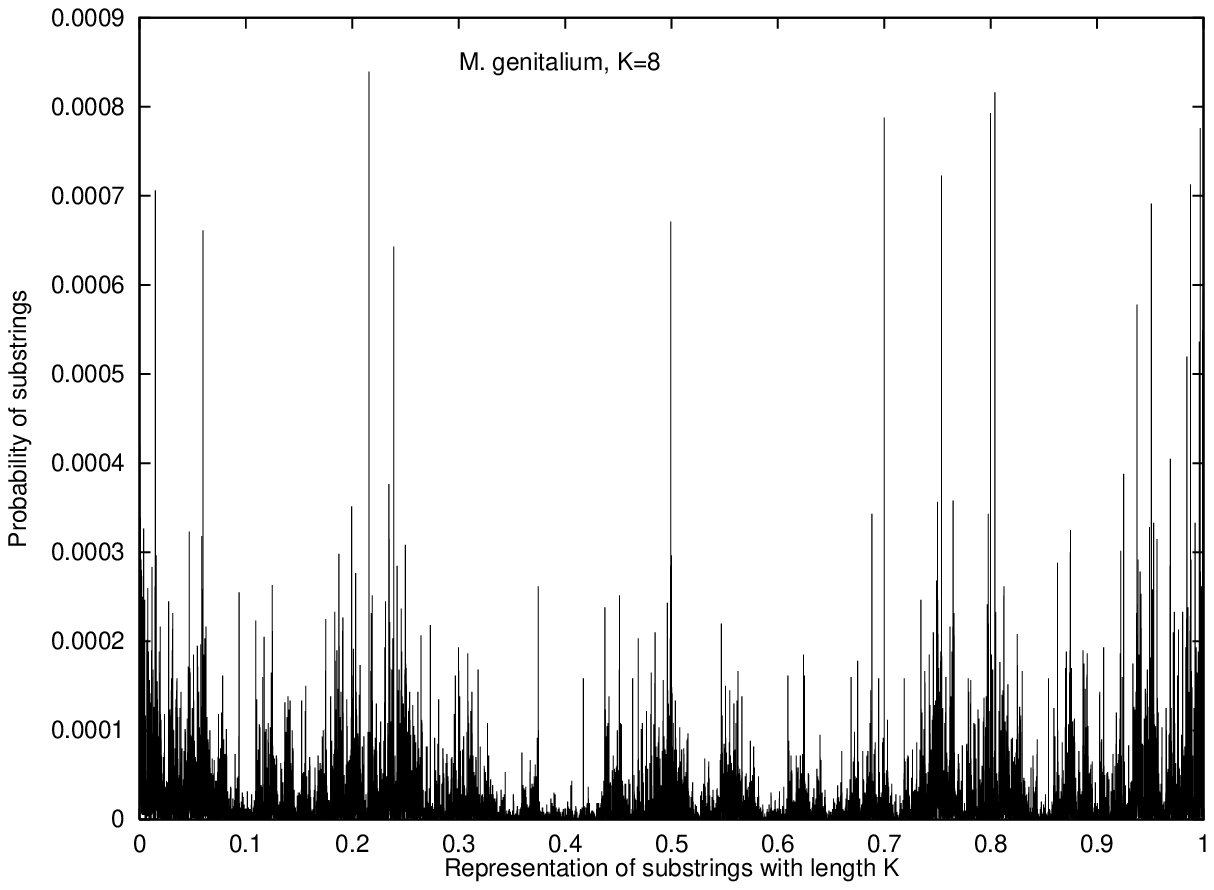}}
\caption{{\protect\footnotesize Histograms  of substrings
with different lengths}}
\label{mgencd}
\end{figure}

We call any string made of $K$ letters from the set $\{g,c,a,t\}$ a $K$%
-string. For a given $K$ there are in total $4^{K}$ different $K$-strings.
In order to count the number of each kind of $K$-strings in a given DNA
sequence $4^{K}$ counters are needed. We divide the interval $[0,1[$ into $%
4^{K}$ disjoint subintervals, and use each subinterval to represent a
counter. Letting $s=s_{1}\cdots s_{K},s_{i}\in \{a,c,g,t\},i=1,\cdots ,K,$
be a substring with length $K$, we define 
\begin{equation}
x_{l}(s)=\sum_{i=1}^{K}\frac{x_{i}}{4^{i}},
\end{equation}
where 
\begin{equation}
x_{i}=\left\{ 
\begin{array}{l}
0,\ \ \ \mbox{if}\ s_{i}=a, \\ 
1,\ \ \ \mbox{if}\ s_{i}=c, \\ 
2,\ \ \ \mbox{if}\ s_{i}=g, \\ 
3,\ \ \ \mbox{if}\ s_{i}=t,
\end{array}
\right. 
\label{order}
\end{equation}
and 
\begin{equation}
x_{r}(s)=x_{l}(s)+\frac{1}{4^{K}}.
\end{equation}
We then use the subinterval $[x_{l}(s),x_{r}(s)[$ to represent substring $s$%
. Let $N_K(s)$ be the number of times that substring $s$ with length $K$ 
appears in the complete genome.
If the number of bases in the complete genome is $L$, we define 
\begin{equation}
F_K(s)=N_K(s)/(L-K+1)
\end{equation}
to be the frequency of substring $s$. It follows that  $\sum_{\{s\}}F_K(s)=1$.
Now we can define a measure $\mu_K$ on $[0,1[$ by $d\mu_K(x)=Y(x)dx$, where
\begin{equation}
Y_K(x)=4^K F_K(s),\ \ \mbox{when}\ \ x\in \lbrack x_{l}(s),x_{r}(s)[.
\end{equation}
It is easy to see $\int_0^1d\mu_K(x)=1$ and $\mu_K([x_l(s),x_r(s)[)=F_K(s)$.
 We call $\mu_K$
the {\it \ measure representation} of the organism corresponding to the given $K$. 
As an example, the
histogram of substrings in the genome  of {\it M. genitalium} 
for $K=3,...,8$ are given  in
FIG. \ref{mgencd}. Self-similarity is apparent in the measure.

For simplicity of notation, the index $K$ is dropped in $F_K(s)$, etc., from
now on, where its meaning is clear. 

\begin{figure}[tbp]
\centerline{\epsfxsize=8cm \epsfbox{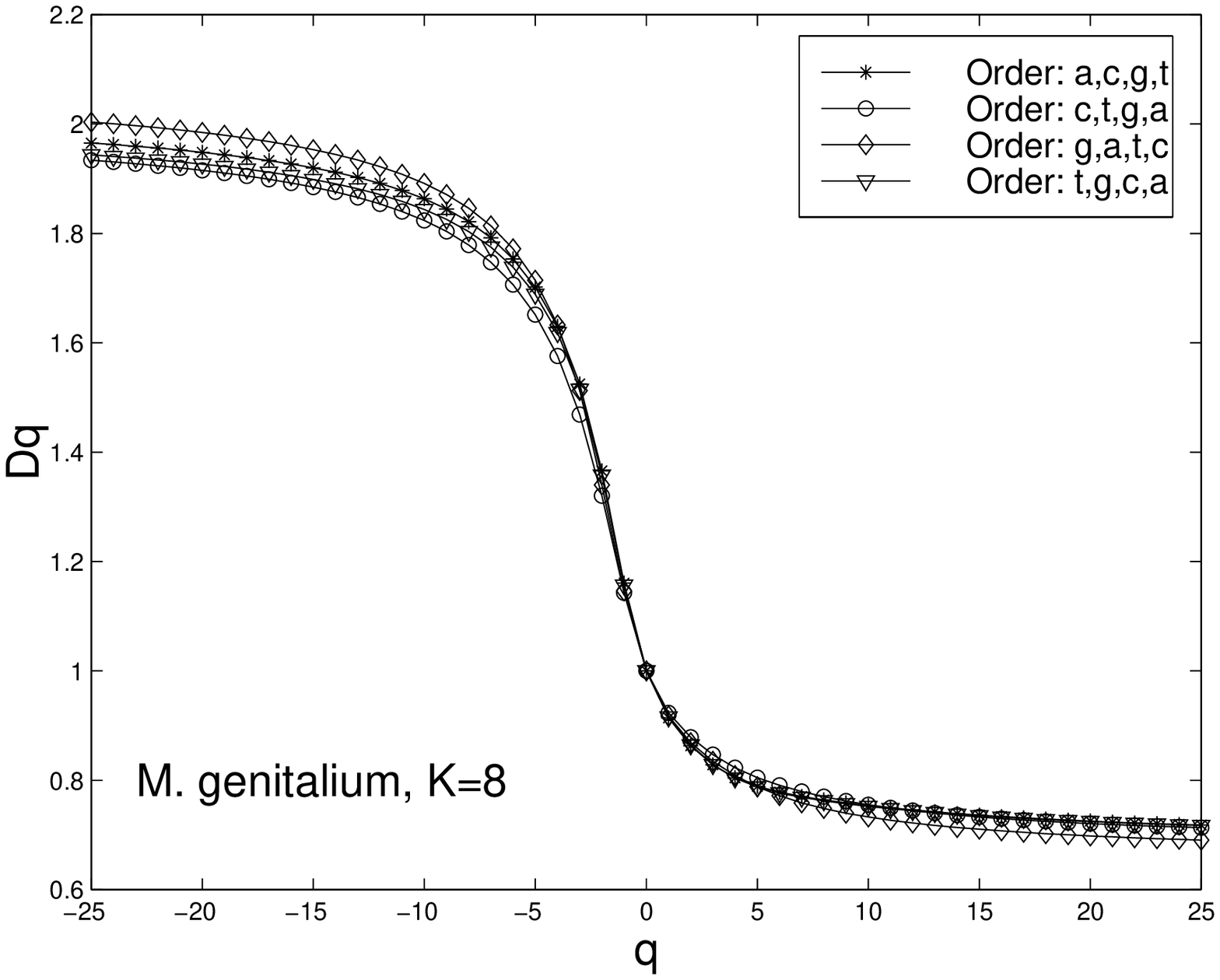}
\epsfxsize=8cm \epsfbox{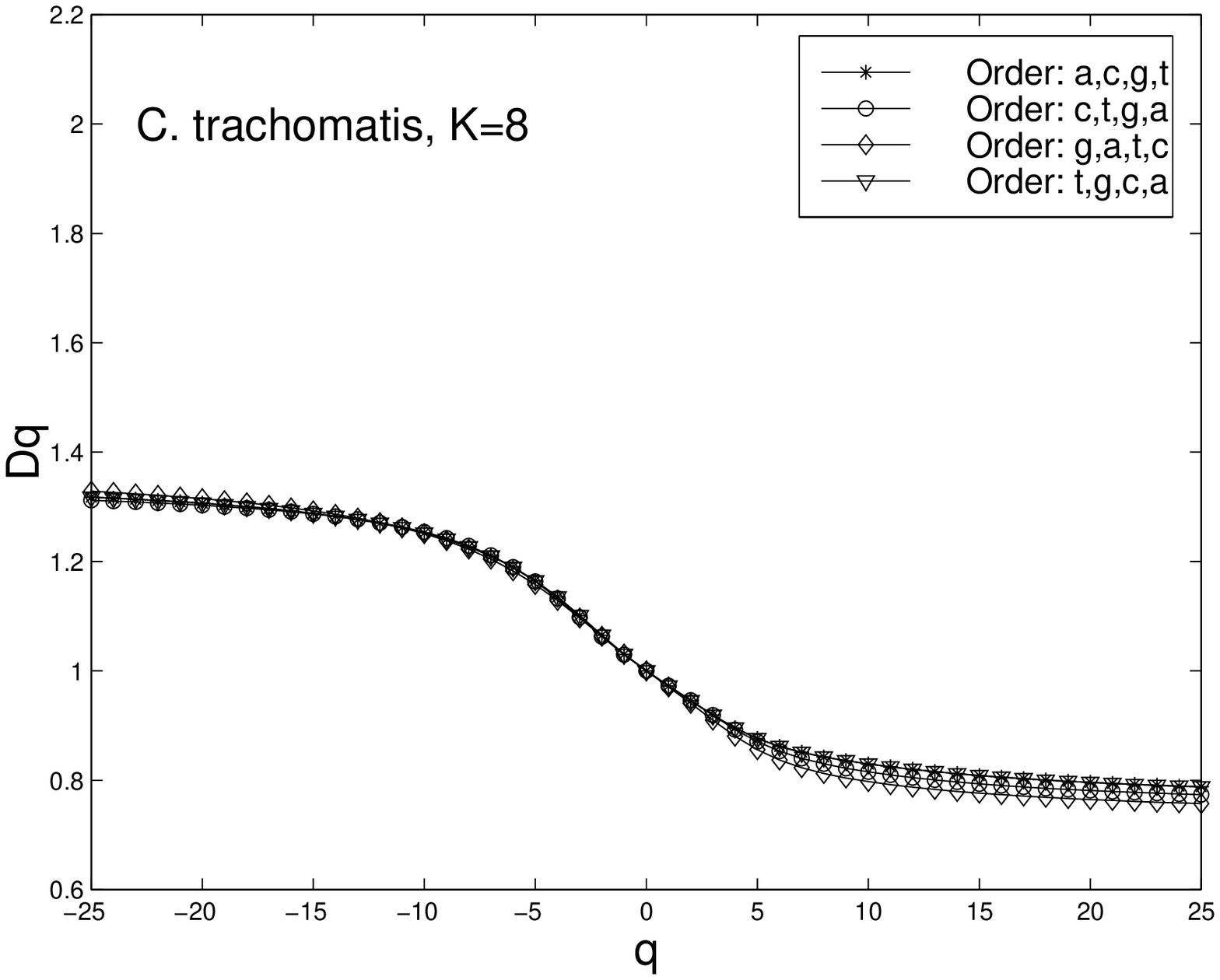}}
\caption{{\protect\footnotesize The dimension spectra of measure representations
given by different ordering of $a,c,g,t$ in (\ref{order}.
 }}
\label{ordercompare}
\end{figure}

  {\bf Remark:} The ordering of $a,c,g,t$ in (\ref{order}) will give the natural
dictionary ordering of $K$-strings in the one-dimensional space. A different ordering 
of $K$-strings would change the nature of the correlations. But in our case, a
different ordering of $a,c,g,t$ in (\ref{order}) give almost tha same $D_q$ curve
(therefore, the same with the $C_q$ curve) which will be defined in the next section when the absolute
value of $q$ is relative small. We give the FIG. \ref{ordercompare} to support
this point of view. Hence a different ordering of $a,c,g,t$ in (\ref{order}) will 
not change our result. When we want to compare different bacteria using the
measure representation, once the ordering of $a,c,g,t$ in (\ref{order}) is given,
it is fixed for all bacteria.

\section{Spectral and multifractal analyses}

We can order all the $F(s)$ according to the increasing order of $x_{l}(s)$.
We then obtain a sequence of real numbers consisting of  $4^{K}$ elements
which we denote as $F(t),t=1,\cdots ,4^{K}$. Viewing the sequence $\{F(t)\}_{t=1}^{4^K}$
as a time series, the spectral analysis can then be undertaken on the sequence.

\begin{figure}[tbp]
\centerline{\epsfxsize=12.5cm \epsfbox{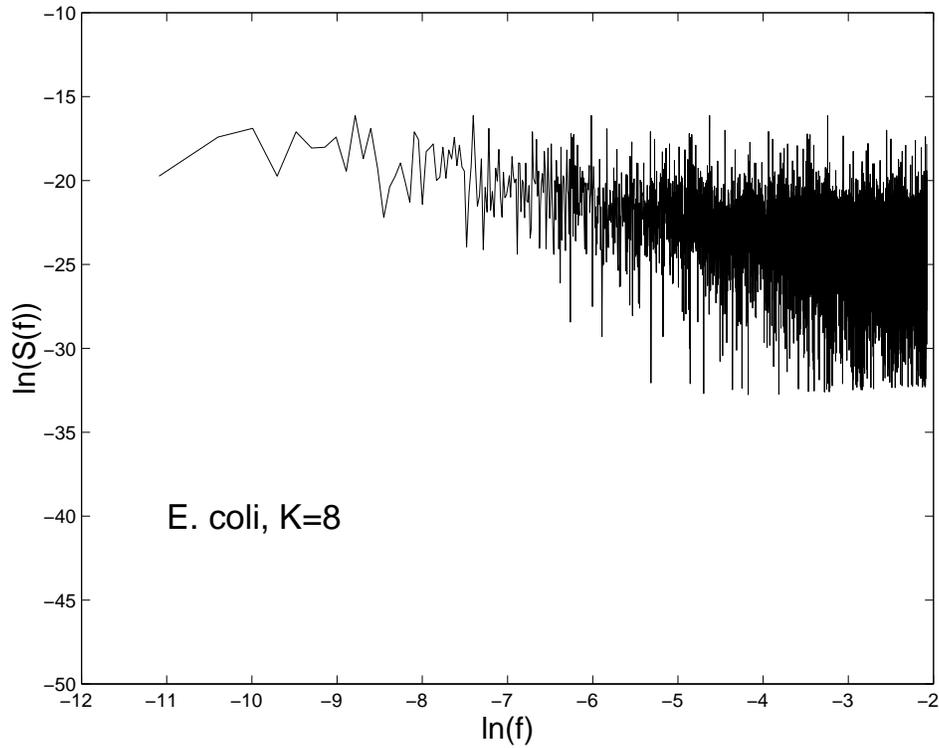}
}
\caption{{\protect\footnotesize The log power spectrum of the measure of E. coli corresponding to
$K=8$. The estimated value of $\beta$ is 0.5986912.}}
\label{ecolipu}
\end{figure}

We first consider the discrete Fourier transform$^{\cite{Shu88}}$ of the
time series $F(t),t=1,\cdots ,4^{K},$ defined by 
\begin{equation}
\widehat{F}(f)=N^{-\frac{1}{2}}\sum_{t=0}^{N-1}F(t+1)e^{-2\pi ift}.
\end{equation}
Then 
\begin{equation}
S(f)=|\widehat{F}(f)|^{2}
\end{equation}
is the {\it power spectrum of } $F\left( t\right) $. In recent studies, it
has been found $^{\cite{Rob74}}$ that many natural phenomena lead to the
power spectrum of the form $1/f^{\beta }$. This kind of dependence was named 
$1/f$ noise, in contrast to white noise $S(f)=const$, i.e. $\beta =0$. Let
the frequency $f$ take $k$ values $f_{k}=k/N,k=1,\cdots ,N/8$. From the $\ln
(S(f))$ vs. $\ln (f)$ graph, we can infer the value of $\beta $ using the
above low-frequency range. For
example, we give the log power spectrum of the measure  of {\it %
E. coli} with $K=8$ in FIG. \ref{ecolipu}.

\begin{figure}[tbp]
\centerline{\epsfxsize=8cm \epsfbox{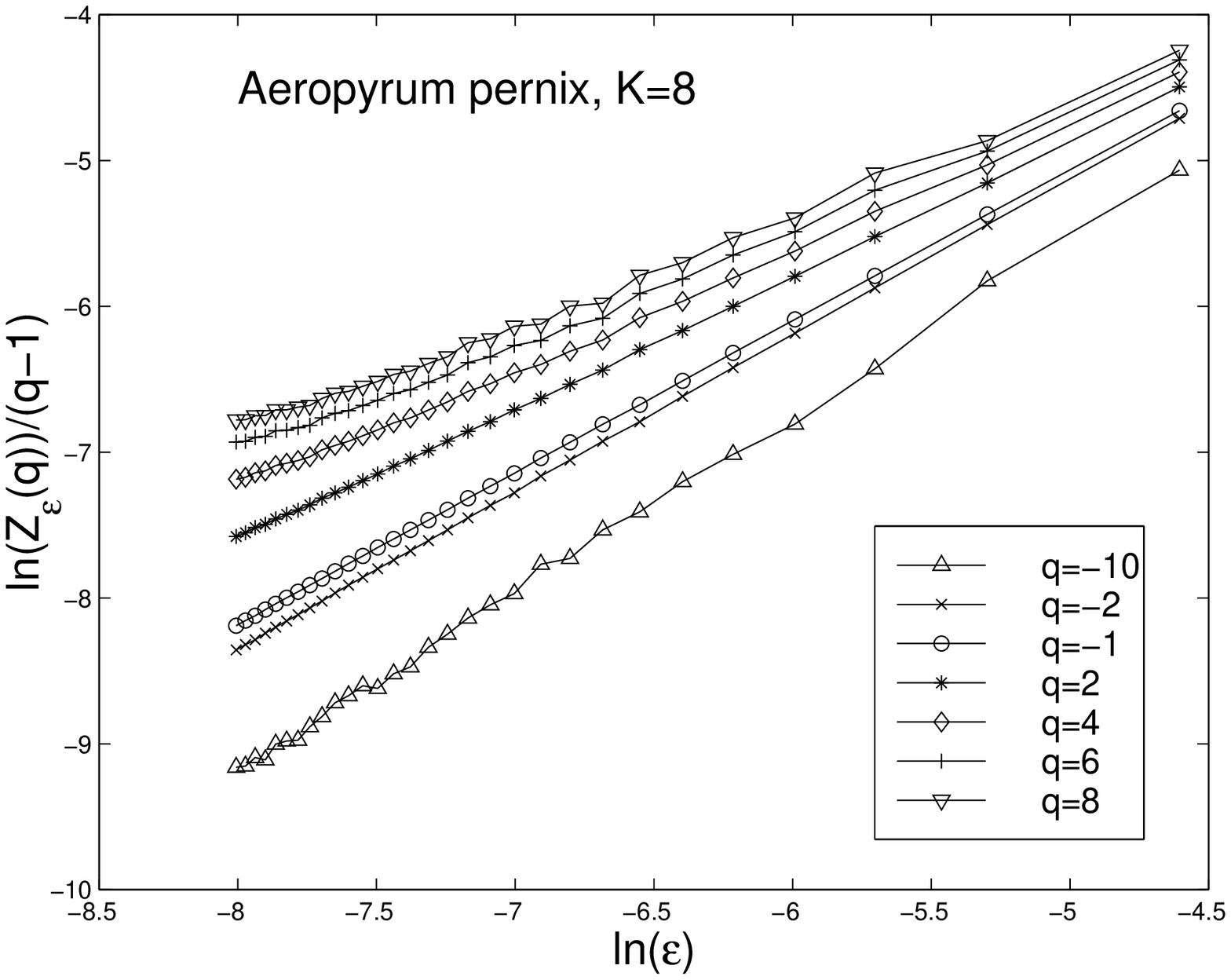}
\epsfxsize=8cm \epsfbox{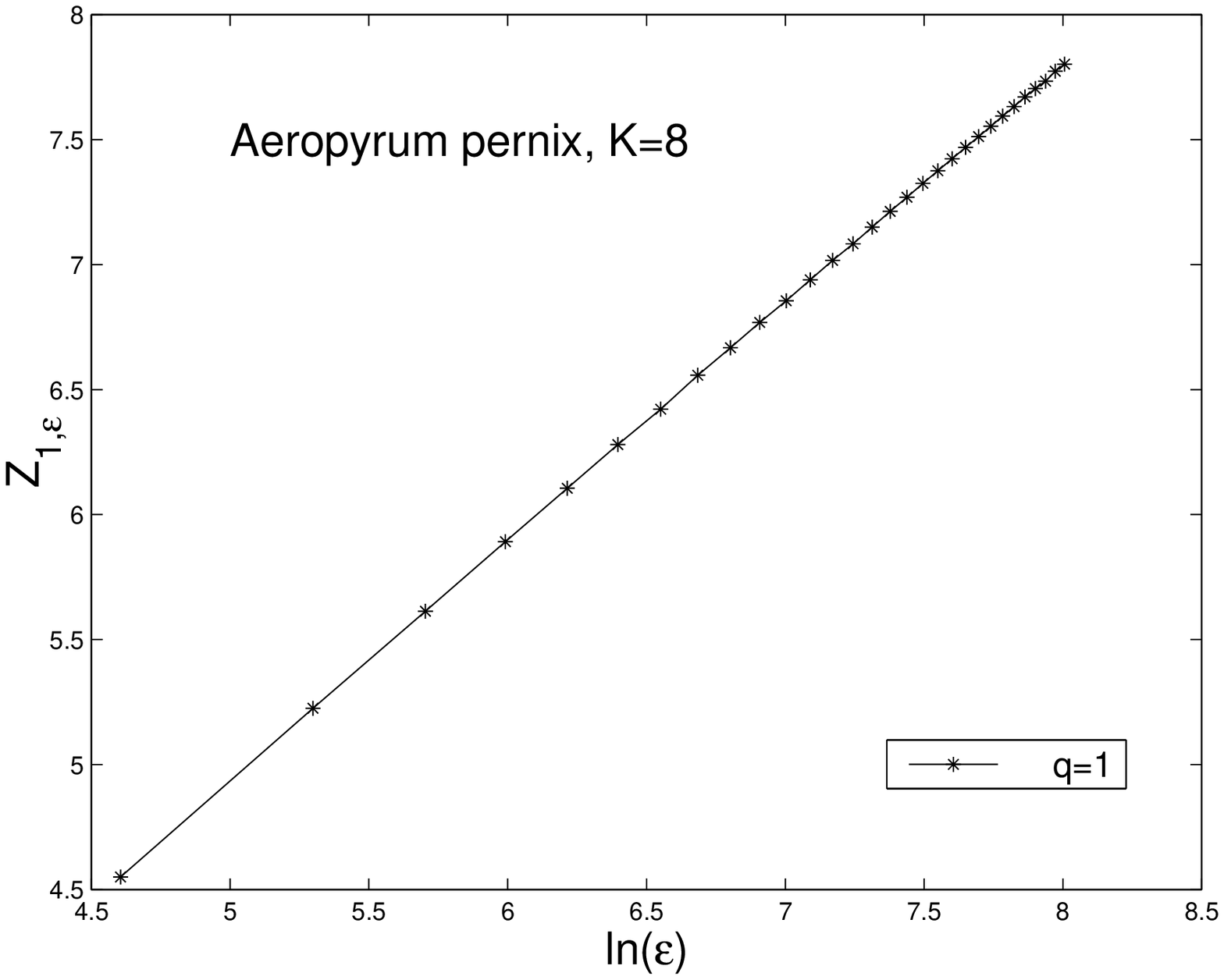}}
\caption{{\protect\footnotesize The linear slopes in the $D_q$
spectra.}}
\label{nihemulti}
\end{figure}

The most common operative numerical implementations of multifractal analysis
are the so-called {\it fixed-size box-counting algorithms} $^{\cite{hjkps}}$%
. In the one-dimensional case, for a given measure $\mu $ with support $%
E\subset {\bf R}$, we consider the {\it partition sum} 
\begin{equation}
Z_{\epsilon }(q)=\sum_{\mu (B)\neq 0}[\mu (B)]^{q},
\end{equation}
$q\in {\bf R}$, where the sum runs over all  different nonempty boxes $B$
of a given side $\epsilon $ in a grid covering of the support $E$, that is, 
\begin{equation}
B=[k\epsilon ,(k+1)\epsilon \lbrack .
\end{equation}
The exponent $\tau (q)$ is defined by 
\begin{equation}
\tau (q)=\lim_{\epsilon \rightarrow 0}\frac{\ln Z_{\epsilon }(q)}{\ln
\epsilon }
\end{equation}
and the generalized fractal dimensions of the measure are defined as 
\begin{equation}
D_{q}=\tau (q)/(q-1),\ \ \mbox{for}\ q\neq 1,
\end{equation}
and 
\begin{equation}
D_{q}=\lim_{\epsilon \rightarrow 0} \frac{Z_{1,\epsilon}}{\ln \epsilon },\ \ \mbox{for}\ q=1.
\end{equation}
where $Z_{1,\epsilon}=\sum_{\mu (B)\neq 0}\mu (B)\ln \mu(B)$.
The generalized fractal dimensions are numerically estimated through a
linear regression of 
\[
\frac{1}{q-1}\ln Z_{\epsilon }(q)
\]
against $\ln \epsilon $ for $q\neq 1$, and similarly through a linear
regression of $Z_{1,\epsilon}$
against $\log \epsilon $ for $q=1$. For example, we show how to obtain the $%
D_{q}$ spectrum using the slope of the linear regression in FIG. \ref
{nihemulti}. $D_1$ is called  {\it information dimension} and $D_2$ is called
 {\it correlation dimension}. The $D_q$ of the positive values of $q$ give relevance
to the regions where the measure is large, i.e., to the $K$-strings with
high probability. The $D_q$ of the negative values of $q$ deal with the
structure and the properties of the most rarefied regions of the measure.

Some sets of physical interest have a nonanalytic dependence of $D_{q}$ on $q
$. Moreover, this phenomenon has a direct analogy to the phenomenon of phase
transitions in condensed-matter physics$^{\cite{KP87}}$. The existence and
type of phase transitions might turn out to be a worthwhile characterisation
of universality classes for the structures$^{\cite{Boj87}}$. The concept of
phase transition in multifractal spectra was introduced in the study of
logistic maps, Julia sets and other simple systems. Evidence of phase transition 
was  found
in the multifractal spectrum of diffusion-limited aggregation$^{\cite{LeS88}}
$. By following the thermodynamic formulation of multifractal measures,
Canessa$^{\cite{can00}}$ derived an expression for the 'analogous' specific
heat as 
\begin{equation}
C_{q}\equiv -\frac{\partial ^{2}\tau (q)}{\partial q^{2}}\approx 2\tau
(q)-\tau (q+1)-\tau (q-1).
\end{equation}
He showed that the form of $C_{q}$ resembles a classical phase transition at
a critical point for financial time series. In the next section, we discuss
the property of $C_{q}$ for our measure representations of organisms.

\section{ Data and results}

 More than 33 bacterial complete genomes are now available in public
databases. There are six Archaebacteria: {\it Archaeoglobus fulgidus}, {\it %
Pyrococcus abyssi}, {\it Methanococcus jannaschii}, {\it Pyrococcus
horikoshii}, {\it Aeropyrum pernix} and {\it Methanobacterium
thermoautotrophicum}; five Gram-positive Eubacteria: {\it Mycobacterium
tuberculosis}, {\it Mycoplasma pneumoniae}, {\it Mycoplasma genitalium}, 
{\it Ureaplasma urealyticum}, and {\it Bacillus subtilis}. The others are
Gram-negative Eubacteria, which consist of two Hyperthermophilic bacteria: 
{\it Aquifex aeolicus} and {\it Thermotoga maritima}; four Chlamydia: {\it %
Chlamydia trachomatisserovar}, {\it Chlamydia muridarum}, {\it Chlamydia
pneumoniae} and {\it Chlamydia pneumoniae AR39}; 
two Spirochaete: {\it Borrelia burgdorferi} and {\it Treponema
pallidum}; one Cyanobacterium: {\it Synechocystis sp. PCC6803}; and thirteen
Proteobacteria. The thirteen Proteobacteria are divided into four
subdivisions, which are alpha subdivision: {\it Rhizobium sp. NGR234} and 
{\it Rickettsia prowazekii}; gamma subdivision: {\it Escherichia coli}, {\it %
Haemophilus influenzae}, {\it Xylella fastidiosa}, {\it Vibrio cholerae}, 
{\it Pseudomonas aeruginosa} and {\it Buchnera sp. APS}; beta subdivision: 
{\it Neisseria meningitidis MC58} and {\it Neisseria meningitidis Z2491};
epsilon subdivision: {\it Helicobacter pylori J99}, {\it Helicobacter pylori
26695} and {\it Campylobacter jejuni}. 

And the complete sequences of some
chromosomes of non-bacteria organisms are also currently available.
In order to discuss the classification problem of bacteria. We also 
selected
the sequences of Chromosome 15 of {\it Saccharomyces cerevisiae}, Chromosome
3 of {\it Plasmodium falciparum}, Chromosome 1 of {\it Caenorhabditis elegans%
}, Chromosome 2 of {\it Arabidopsis thaliana} and Chromosome 22 of {\it Homo
sapiens}.

\begin{figure}[tbp]
\centerline{\epsfxsize=12.5cm \epsfbox{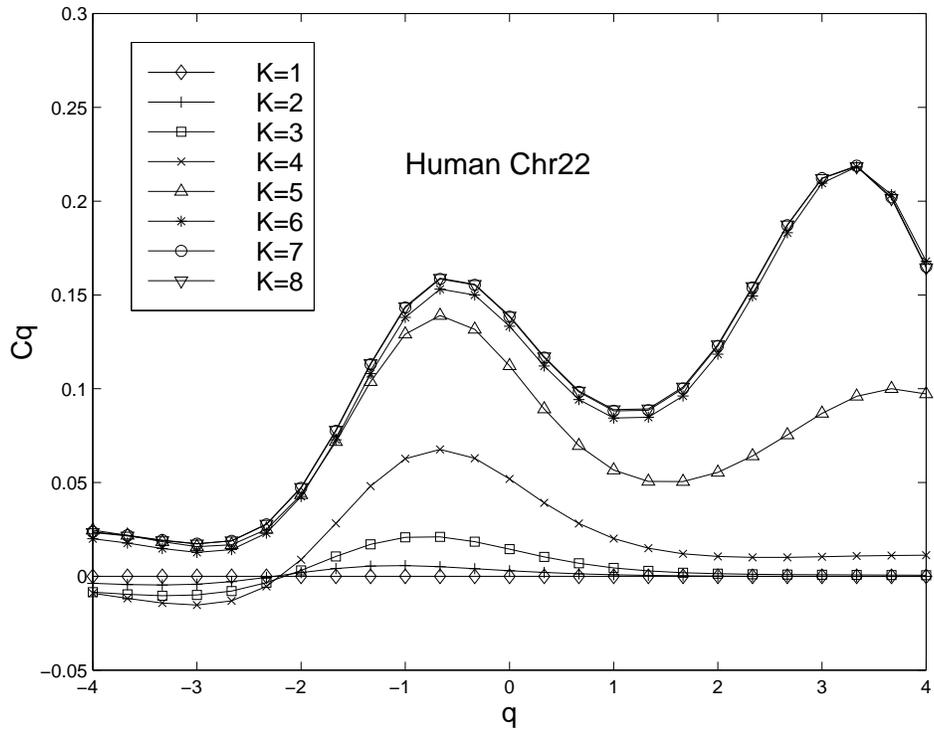}
}
\caption{{\protect\footnotesize "Analogous" specific heat of
measures of substrings with different lengths $K$ in Chromosome 22 of
Homo sapiens.}}
\label{Cqk}
\end{figure}

\begin{figure}[tbp]
\centerline{\epsfxsize=11.5cm \epsfbox{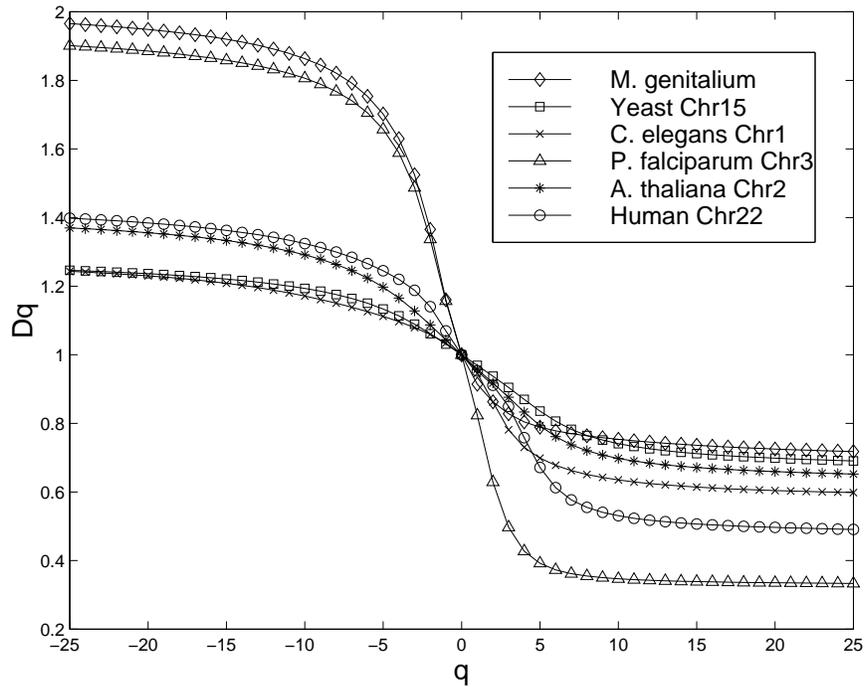}
}
\caption{{\protect\footnotesize Dimension spectra of Chromosome 22 of Homo sapiens,
Chromosome 2 of A. thaliana, Chromosome 3 of P. falciparum, Chromosome 1 of
C. elegans, Chromosome 15 of S. cerevisiae and M. genitalium. }}
\label{Dqjinhua}
\end{figure}

\begin{figure}[tbp]
\centerline{\epsfxsize=11.5cm \epsfbox{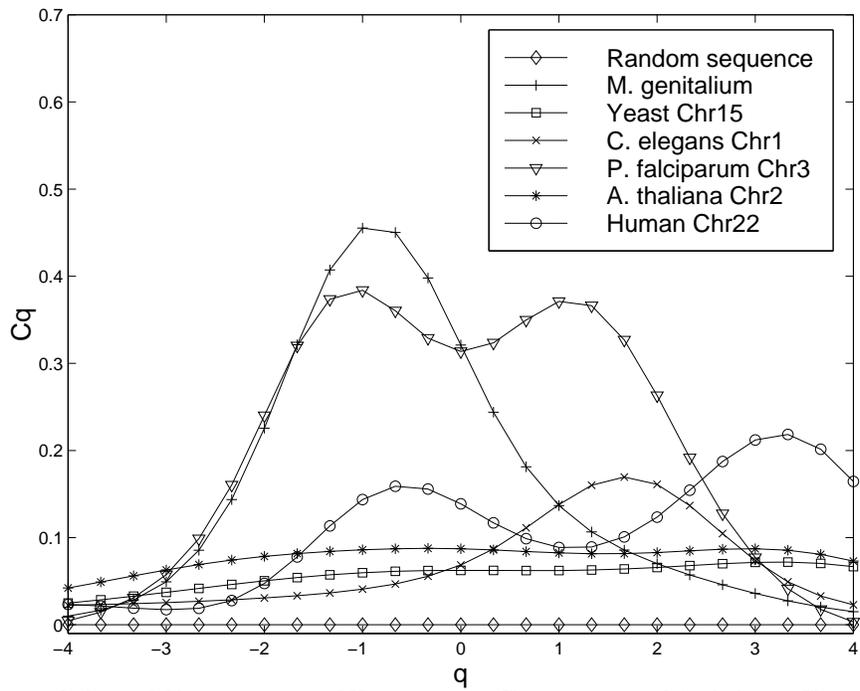}
}
\caption{{\protect\footnotesize "Analogous" specific heat of Chromosome 22
of Homo sapiens, Chromosome 2 of A. thaliana, Chromosome 3 of P. falciparum,
Chromosome 1 of C. elegans, Chromosome 15 of S. cerevisiae, M. genitalium and
complete random sequence. }}
\label{Cqjinhua}
\end{figure}

We obtained the dimension spectra and 'analogous' specific heat of the
measure representations of the above organisms and used them to discuss the
classification problem. We calculated the dimension spectra
and 'analogous' specific heat of chromosome 22 of Homo sapiens for $K=1,...,$%
8, and found that the $D_{q}$ and $C_{q}$ curves of $K=6,7,8$ are very close
to one another (see FIG. \ref{Dqk} and \ref{Cqk}). Hence it seems
appropriate to use the measure corresponding to $K=8$. For $K=8$, 
we calculated the dimension spectra,
 'analogous' specific heat and the
exponent $\beta$ of the measure representations 
of all the above organisms. As an illustration, we plot the $D_{q}$ curves
of {\it M. genitalium}, Chromosome 15 of {\it %
Saccharomyces cerevisiae}, Chromosome 3 of {\it Plasmodium falciparum},
Chromosome 2 of {\it Arabidopsis thaliana} and Chromosome 22 of {\it Homo
sapiens} in FIG. \ref{Dqjinhua}; and the $C_{q}$ curves of these organisms
in FIG. \ref{Cqjinhua}. Because all $D_q$ are equal to 1 for the complete random sequence,
from these plots, it is apparent that the $D_{q}$
and $C_{q}$ curves are nonlinear and significantly different from those of
the completely random sequence.  From FIG. \ref{Dqjinhua}, we can claim that
the curves representative of the organisms are clearly distinct from
the curve representing a random sequence.
From the plot of $D_{q}$, the dimension spectra
of organisms exhibit a multifractal-like form. From FIG. \ref{nihemulti}, we can
see the linear fits of $q=-2,-1,1,2$ are perfect and better than that
of other values of $q$,
Hence we suggest to use $D_{-2},D_{-1},D_1,D_2$ in the
comparison of different bacteria. We give the numerical results for 
$D_{-2},D_{-1},D_1,D_2$ in
Table \ref{tabled25} (from top to bottom, in the increasing order of the
value of $D_{-1}$).

\begin{figure}[tbp]
\centerline{\epsfxsize=8cm \epsfbox{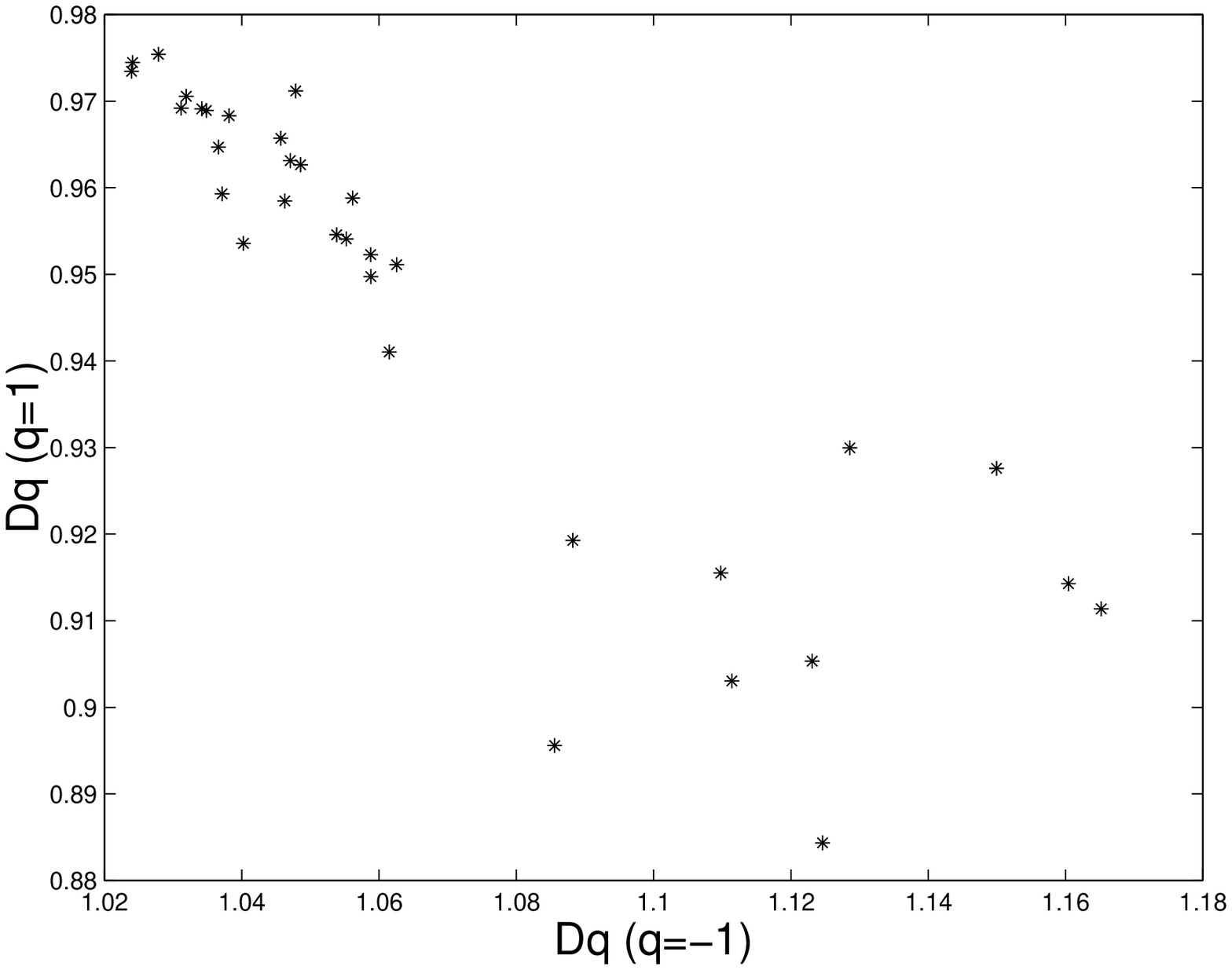}
\epsfxsize=8cm \epsfbox{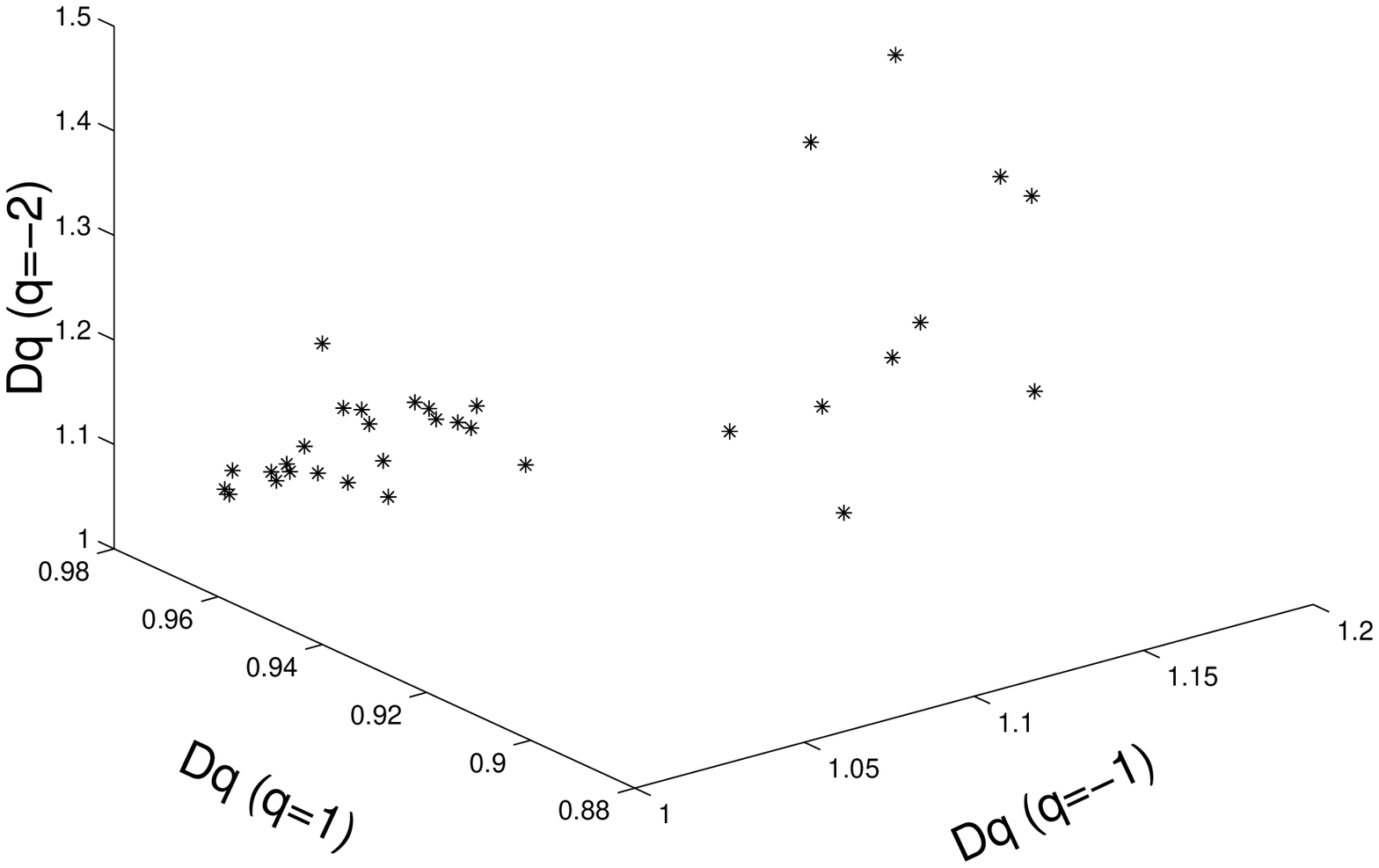}}
\caption{{\protect\footnotesize The distributions of two-dimensional points ($\protect%
D_{-1}$,$D_{1}$) and three-dimensional points ($\protect%
D_{-1}$,$D_{1}$,$D_{-2}$) of the bacteria selected. }}
\label{twobetadq}
\end{figure}

If only a few bacteria are considered at a time, we can use the $D_{q}$
curve to distinguish them. This strategy is clearly not efficient when a
large number of organisms are to be distinguished. For this purpose, we
suggest to use $D_{-1},D_1$ and $D_{-2}$, in conjunction with 
two-dimensional points $(D_{-1} ,D_{1})$ or three-dimensional points 
$(D_{-1},D_1,D_{-2})$. We give the distribution of
two-dimensional points $(D_{-1},D_{1})$ and three-dimensional points
$(D_{-1},D_1,D_{-2})$ of bacteria in FIG. \ref{twobetadq}. 

\section{Discussion and conclusions}

The idea of our measure representation is similar to the portrait method
proposed by Hao {\it et al.}$^{\cite{hlz98}}$. It provides a simple yet
powerful visualisation method to amplify the difference between a DNA
sequence and a random sequence as well as to distinguish DNA sequences
themselves in more details. If a DNA sequence is random, then our measure
representation yields a uniform measure ($D_{q}=1,\ C_{q}=0$).

From the measure representation and the values of $D_{q}$ and $C_{q}$, it is
seen that there exists a clear difference between the DNA sequences of all
organisms considered here and the completely random sequence. Hence we can
conclude that complete genomes are not random sequences.

We obtained the values of the exponent $\beta $ of our measure representations 
($\beta=0.393003$ for {\it V. cholerae}, $\beta=0.311623$ for {\it A. pernix},
$\beta=0.240601$ for {\it X. fastidiosa}, $\beta=0.381293$ for {\it T. pallidum},
$\beta=0.334057$ for {\it C. pneumoniae AR39}, and $\beta$ is larger than
0.4 for all other bacteria selected).  These values are far from $0$.
Hence when we view our measure representations of organisms as time series,
they are far from being random time series, and in fact exhibit strong long-range
correlation. Here the long-range correlation is for the $K$-strings
with the dictionary ordering, and it is different from the base pair correlations
introduced by other people.

Although the existence of the archaebacterial urkingdom has been accepted by
many biologists, the classification of bacteria is still a matter of
controversy$^{\cite{iwabe}}$. The evolutionary relationship of the three
primary kingdoms, namely archeabacteria, eubacteria and eukaryote, is
another crucial problem that remains unresolved$^{\cite{iwabe}}$.

\begin{figure}[tbp]
\centerline{\epsfxsize=12.5cm \epsfbox{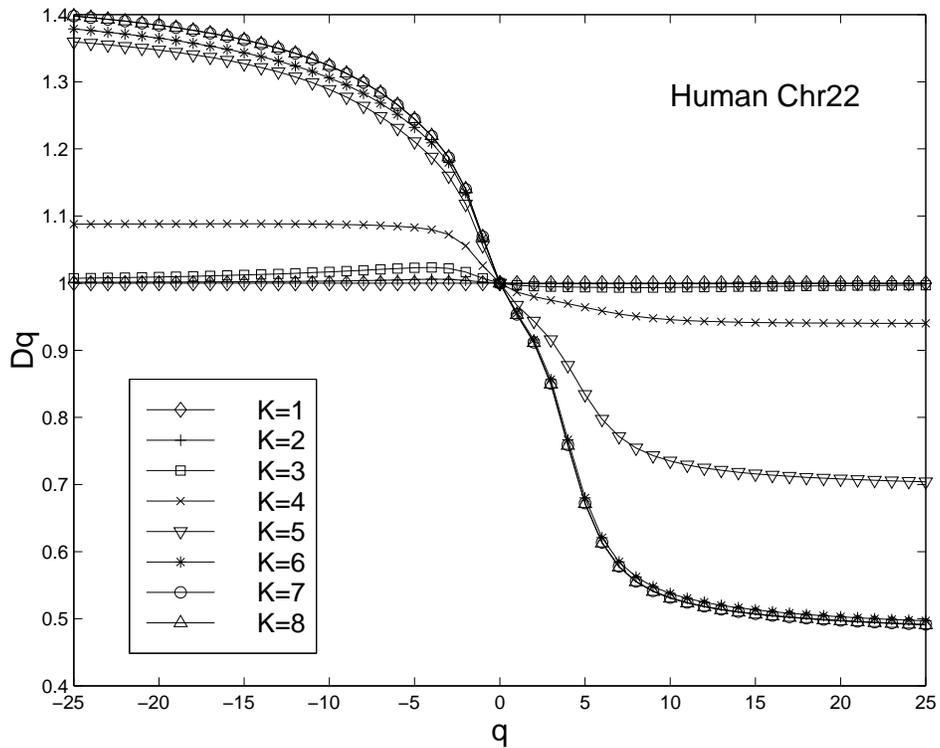}
}
\caption{{\protect\footnotesize Dimension spectra of measures of
substrings with different lengths $K$ in Chromosome 22 of Homo sapiens. }}
\label{Dqk}
\end{figure}

When $K$ is large ($K\geq 6$), our measure representation contains rich
information on the complete genomes. From FIG. \ref{Dqk} and FIG. \ref{Cqk},
we find the curves of $D_{q}$ and $C_{q}$ are very close to one another for $%
K=6,7,8$. Hence, for the classification problem, it would be
appropriate to take $K=8$. We calculated the $\beta $, $D_{q}$ and $C_{q}$
values of all organisms selected in this paper for $K=8$. We found that the $%
D_{q}$ spectra of all organisms are multifractal-like and sufficiently
smooth so that the $C_{q}$ curves can be meaningfully estimated. From FIG. 
\ref{Dqk}, with the decreasing of $K$, the multifractality becomes less
severe. 
With $K=8$, we found that the $C_{q}$ curves of all other bacteria resemble
a classical phase transition at a critical point similar to that of {\it M.
genitalium} shown in FIG. \ref{Cqjinhua}. But the 'analogous' phase
transitions of non-bacteria organisms are different. Apart from Chromosome 1 of  
{\it C. elegans, }they exhibit the shape of double-peaked specific heat
function which is known to appear in the Hubbard model within the {\it %
weak-to-strong} coupling regime$^{\cite{vol97}}$.

It is seen that the $D_{q}$ curve is not clear enough  to distinguish many
bacteria themselves. In order to solve this problem we  use
 two-dimensional points $(D_{-1} ,D_{1})$ and three-dimensional points 
$(D_{-1},D_1,D_{-2})$. From FIG.%
\ref{twobetadq}, it is clear that bacteria roughly gather into two classes
(as shown in Table \ref{tabled25}).  Using the
distance among the points, one can obtain a classification of bacteria. 

From Table \ref{tabled25}, we can see all Archaebacteria belong to the same class
except {\it M. jannaschii}. And four Chlamydia almost gather together. 
It is surprised that the closest pairs of bacteria, {\it Helicobacter pylori J99} 
and {\it %
Helicobacter pylori 26695}, {\it Neisseria meningitidis MC58} and 
{\it Neisseria meningitidis Z2491},  group with each other. Two hyperthermophilic bacteria
group with each other and are linked with the Archaebacteria.
It has previously been shown that 
{\it Aquifex} has close relationship with Archaebacteria from the gene comparison of
 an enzyme needed for the synthesis of the amino acid 
 trytophan$^{\cite{pennisi}}$ and using the length sequence of complete 
 genome$^{\cite{YA00}}$. 
In general, Bacteria that are
close phylogenetically are almost close in the spaces $(D_{-1},D_1)$
and $(D_{-1},D_1,D_{-2})$.

\section*{Acknowledgement}

\ \ \ One of the authors, Zu-Guo Yu, would like to express his gratitude to
Prof. Bai-lin Hao of Institute of Theoretical Physics of the Chinese Academy
of Science for introducing him into this field and continuous encouragement.
He also wants to thank Dr. Enrique Canessa of ICTP for pointing out the
importance of the quantity $C_{q}$ and useful comments, and Dr. Guo-Yi Chen
of ITP for useful suggestions on the measure representation. In particular,
he want to thanks one of the referee for suggestion using the $D_q$ of
negative values of $q$ to classify the bacteria. This research
was partially supported by QUT Postdoctoral Research Support Grant 9900658,
and the HKRGC Earmarked Grant CUHK 4215/99P.

\onecolumn
\begin{table}
\caption{The values of $D_{-1},D_1,D_{-2}$ and $D_2$ of all
bacteria selected.}
\label{tabled25}
\begin{center}
\begin{tabular}{|l|l|c|c|c|c|}
\ \ \ \ \ \ Species & \ \ \ \ \ \ Category &  $D_{-1}$ & $%
D_1 $ & $D_{-2}$ & $D_2$ \\ \hline
 Xylella fastidiosa & Proteobacteria &  1.023935 &     0.9734505  &     1.046237   &   0.9434007 \\
 Treponema pallidum & Spirochaete &  1.024096  &    0.9744529  &     1.048537  &    0.9456879 \\   
 Vibrio cholerae & Proteobacteria &  1.027849  &    0.9754193  &     1.060974  &    0.9529402 \\   
 Bacillus subtilis & Gram-positive Eubacteria &  1.031173  &    0.9691831   &    1.062364  &    0.9392986 \\
 Chlamydia trachomatis & Chlamydia & 1.031900  &    0.9705723   &    1.067158  &    0.9421241 \\   
 Chlamydia pneumoniae & Chlamydia &  1.034190  &    0.9691189   &    1.075935  &    0.9396138 \\   
 Rhizobium sp. NGR234 & Proteobacteria &  1.034821  &    0.9689233  &     1.068532  &    0.9430141 \\   
 Chlamydia muridarum & Chlamydia &  1.036608  &    0.9646960   &    1.075166  &    0.9293640\\    
 Chlamydia pneumoniae AR39 & Chlamydia &  1.037127  &    0.9593074   &    1.078164   &   0.9106171\\    
 Pyrococcus abyssi & Archaebacteria &  1.038142  &    0.9683081   &    1.091387  &    0.9393384 \\
 Aeropyrum pernix & Archaebacteria &  1.040248   &   0.9535630   &    1.074807   &   0.9033159 \\   
 Synechocystis sp. PCC6803 & Cyanobacteria &  1.045674  &    0.9657137   &    1.127265  &    0.9364141 \\   
 Mycoplasma pneumoniae & Gram-positive Eubacteria &  1.046260   &   0.9584649   &    1.092869  & 0.9250106\\ 
 Archaeoglobus fulgidus & Archaebacteria &  1.047071  &    0.9631252  &     1.130371  &    0.9279480 \\  
 Escherichia coli & Proteobacteria &  1.047849   &   0.9711645   &    1.174754  &    0.9474317 \\   
 M. thermoautotrophicum & Archaebacteria &  1.048569  &    0.9626480  &     1.116451  &    0.9306760\\    
 Thermotoga maritima & Hyperthermophilic bacteria &  1.053824   &   0.9545637  &  1.145209  & 0.9101596 \\  
 Aquifex aeolicus & Hyperthermophilic bacteria &  1.055210  &    0.9540893 & 1.134702 &     0.9145361 \\   
 Pyrococcus horikoshii & Archaebacteria &  1.056144   &   0.9587924   &    1.139402  &    0.9237674 \\  
 Neisseria meningitidis MC58 & Proteobacteria &  1.058779  &    0.9522681  & 1.132902 &  0.9132383 \\   
 Neisseria meningitidis Z2491 & Proteobacteria &  1.058805 &     0.9497503  &     1.133201  & 0.9065167 \\   
 M. tuberculosis & Gram-positive Eubacteria &  1.061496  &    0.9410341  & 1.115466  &    0.8920540\\    
 Haemophilus influenzae & Proteobacteria &  1.062565  &    0.9511231  &     1.147970 & 0.9122260 \\   
 \hline\hline
 Buchnera sp. APS & Proteobacteria &  1.085581  &    0.8955851 &      1.152650  &    0.7904221\\    
 Rickettsia prowazekii & Proteobacteria &  1.088237  &    0.9192655  &     1.173883  &    0.8567044 \\   
 Pseudomonas aeruginosa & Proteobacteria &  1.109776  &    0.9154980  &     1.187378  &    0.8622321 \\   
 Borrelia burgdorferi & Spirochaete &  1.111380  &    0.9030539  &     1.261299  &    0.8298323 \\   
 Campylobacter jejuni & Proteobacteria &  1.123096  &    0.9053437  &     1.279505  &    0.8349793 \\   
 Ureaplasma urealyticum & Gram-positive bacteria &  1.124616 &     0.8843481  &     1.260287 & 0.8065916\\
 Helicobacter pylori J99 & Proteobacteria &  1.128590 &     0.9299614    &   1.390791  &    0.8758443 \\   
 Helicobacter pylori 26695 & Proteobacteria &  1.149943  &    0.9276062  &     1.460757  &    0.8719445 \\   
 Mycoplasma genitalium & Gram-positive Eubacteria &  1.160435  & 0.9142718  & 1.365716  & 0.8631789 \\   
 Methanococcus jannaschii & Archaebacteria &  1.165208   &   0.9113731   &    1.349664   &   0.8628226 \\
\end{tabular}
\end{center}
\end{table}

\end{document}